\documentstyle[12pt,aaspp4]{article}
\begin{document}
\title{$\it{FIRST}$ Bent-Double Radio Sources: Tracers of High-Redshift Clusters\footnote{Based in part on observations obtained at the
W. M. Keck Observatory}}
\author{E. L. Blanton\altaffilmark{2,6}, M. D. Gregg\altaffilmark{3,4}, D. J. Helfand\altaffilmark{2,6}, 
R. H. Becker\altaffilmark{3,4}, and R. L. White\altaffilmark{5}}
\altaffiltext{2}{Department of Astronomy, Columbia University, New York, NY  10027}
\altaffiltext{3}{Institute of Geophysics \& Planetary Physics, Lawrence Livermore
National Laboratory, Livermore, CA  94550}
\altaffiltext{4}{Physics Department, University of California, Davis, CA  95616}
\altaffiltext{5}{Space Telescope Science Institute, Baltimore, MD  21218}
\altaffiltext{6}{Visiting Astronomer, Kitt Peak National Observatory, 
National Optical Astronomy Observatories, which is operated by the Association
of Universities for Research in Astronomy, Inc., under cooperative agreement
with the National Science Foundation.}

\begin{abstract}
Bent-double radio sources can act as tracers for clusters of galaxies.  We present
imaging and spectroscopic observations of the environments surrounding ten of 
these sources (most of them Wide Angle Tails (WATs)) selected from the VLA $\it{FIRST}$ survey.  
Our results reveal a previously unknown cluster associated with eight of the radio 
sources with redshifts in the range $0.33 < z < 0.85$; furthermore,we cannot rule
out that the other two bent doubles may be associated with clusters at
higher redshift.  Richness measurements indicate that these clusters are typical
of the majority of those found in the Abell (1958) catalog, with a range of Abell
richness classes from 0 to 2.  The line-of-sight velocity dispersions are very
different from cluster to cluster, ranging from approximately 300 to 1100
km s$^{-1}$.  At the upper end of these intervals, we may be sampling some
of the highest-redshift massive clusters known.  Alternatively, the large
velocity dispersions measured in some of the clusters may indicate that
they are merging systems with significant substructure, consistent with
recent ideas concerning WAT formation (Burns et al. 1994).
\end{abstract}


\section{Introduction}
Clusters of galaxies are the largest gravitationally bound structures in
the universe and provide well-equipped laboratories for the study of galaxy formation and evolution,
as well as a testing ground for the estimation of cosmological parameters.  Thousands of clusters
have been identified at redshifts $z < 0.25$ but for $z > 0.5$,
the number of spectroscopically confirmed clusters dwindles to less than
approximately one hundred.
With $\sim$ 30,000
galaxies deg$^{-2}$ at 24$^{th}$ magnitude in the V-band and fewer than 50 clusters
deg$^{-2}$ at this limit out to a redshift of 1.0 (Postman
1993), identification of high-redshift clusters from optical images of the sky is a 
difficult task.  Nonetheless, most known clusters,
including those with $z > 0.5$, have been discovered in optical surveys
(Abell 1958; Zwicky et al. 1968; Gunn, Hoessel \& Oke 1986; Couch et al. 1991;
Postman et al. 1996).  The clusters detected in the earlier surveys were
selected visually, while Postman et al. (1996) 
used an automated technique for the Palomar Distant Cluster Survey (PDCS) in
which clusters were detected
as simultaneous over-densities in positional and brightness distributions.
Both types of optical surveys suffer from misclassifications due to the
superposition of foreground and background galaxies, although the PDCS technique
significantly reduces this effect.  
At longer wavelengths,
searching in the near-IR reveals high$-z$ clusters as overdensities of very red
galaxies:  Stanford et al. (1997) used this method to locate one of
the most distant clusters known with
several spectroscopically confirmed members at $z = 1.27$.

Other groups have found distant clusters by searching in the X-ray band,
detecting directly the diffuse hot gas which dominates the visible matter in 
most clusters.  The main
drawback to this approach is the limited sensitivity of large-area X-ray surveys, especially given
the possibility that distant clusters are intrinsically dimmer than nearer ones
(Henry et al. 1992; Gioia et al. 1990a), although evolution in the X-ray
luminosity function out to $z \sim 0.8$ is far from established (Rosati et al.
1998).  Only six of the 104 clusters in the
Einstein Extended Medium Sensitivity Survey (EMSS) have $z > 0.5$ (Luppino \&
Gioia 1995;
Stocke
et al. 1991; Gioia et al. 1990b).  The ROSAT all-sky survey has not uncovered
many $z > 0.5$ clusters and several groups have resorted to conducting  
serendipitous surveys of deeper ROSAT pointings.  For example, Rosati et al. (1998)
find evidence for 95 clusters out to $z \sim 0.8$ ($\sim 20$ with $z > 0.5$) in a
48 deg$^{2}$ area using a wavelet-based detection algorithm.  
Other ROSAT-based surveys -- including WARPS (Wide-Angle ROSAT Pointed x-ray 
Survey, Scharf et al. 1997), SHARC (Serendipitous High-Redshift Archival
ROSAT Cluster survey, Burke et al. 1997, Romer et al. 1999), and the survey by Vikhlinin et al. 
(1998) -- are finding several clusters 
beyond $z = 0.5$ in addition to many lower redshift systems.

With increasing redshift, optical and X-ray surveys become
less effective and a targeted approach to finding clusters is warranted.
Another technique for finding clusters is to examine the environments of
radio galaxies.
Dickinson (1996) has identified clusters around powerful 3C sources
at redshifts greater than 1.2. In a similar manner, Deltorn et al. (1997) discovered a cluster of
galaxies around 3CR 184 at $z = 0.996$.  Clusters of varying richness have
also been seen around the radio galaxies studied by Hill \& Lilly (1991 -- 
hereafter H\&L91) and by
Zirbel (1997 -- hereafter Z97), among others.  However, certain types of radio galaxies are more often associated with clusters than others, and are therefore more efficient tracers of 
these high-density environments.

Over two decades ago, Fanaroff \& Riley (1974) discovered a correlation
between the power of an extended radio source and its
morphology and they divided all such sources into two classes.  
FR I radio galaxies
typically have powers less than $5\times10^{25}~\rm{W~Hz^{-1}}$ at 1440 MHz
(assuming a power law spectrum with $\alpha = 0.8$)
and are brightest at their cores, with their lobes fading toward the edges.  
FR II's have $\rm{P_{1440}>5\times10^{25}~W~Hz^{-1}}$ with dim or absent 
cores and edge-brightened lobes.  More recent work has shown that the dividing
line in power between FR I's and II's isn't so sharp, with examples of each
crossing the line (Ledlow \& Owen 1996).  In addition, Ledlow \& Owen (1996) 
found that the
break in radio power is a function of optical brightness with higher power FR I's
hosted by optically brighter galaxies.
 
Recent work (H\&L91, Z97) has shown that FR I and II radio 
galaxies
are associated with different types of host galaxies located in different
environments on the megaparsec scale.  FR I sources are associated with 
elliptical galaxies with a wide range of optical magnitudes (Ledlow \& Owen
1995).  Powerful FR I sources (those near the FR I/II break such as the WATs and
NATs discussed below) are often associated
with cD or double nuclei elliptical galaxies found in rich groups or clusters
at both low and high redshift (up to z $\approx$ 0.5, the limit
of current studies).  FR II sources are usually associated with N galaxies or
disturbed ellipticals and are found in poor groups at low-z; at higher
redshifts ($z \approx 0.5$), groups surrounding FR II's appear to
be richer by a factor of $\sim 2$.
 
A source's radio morphology may become distorted as a consequence of 
the relative motion between the host
galaxy and the surrounding intracluster medium (ICM) or via interaction with a neighboring galaxy.  Common
distortions in FR I sources are wide angle tail (WAT) and narrow angle tail
(NAT) morphologies which have large and small opening angles, respectively,
and are caused by interaction with the ICM.
The classical explanation for
the bending of the lobes is the ram pressure exerted by the ICM as a galaxy
with a significant peculiar velocity moves through a cluster (Owen \& Rudnick 1976).  An alternate
explanation (Burns et al. 1996; Roettiger, Burns, \& Loken 1996; Burns et al. 1993) is that it is the ICM
that is moving and not the galaxy, since WATs are often associated with cD
galaxies which may have small peculiar velocities relative to the immediately
surrounding cluster.  In this scenario, the
ICM is set in motion by the merging of clusters or smaller sub-clumps
evidenced by the alignment of X-ray emission contours with the bending of the lobes
(G\'omez et al. 1997), and the cD then has a larger peculiar velocity relative
to the merging system as a whole.  FR II sources may also become 
distorted; sometimes the distortions appear symmetric and may result from ICM
interaction, whereas at other times the sources have asymmetric 'dog-leg' 
appearances
that are likely caused by interaction with a neighboring galaxy (Rector, Stocke, \&
Ellingson 1995).
 
Since FR I galaxies are located in groups or clusters at all epochs explored to
date, they are markers of the high-density Universe.  Bent (WAT or NAT)
FR I radio galaxies may be signposts for rich clusters (rather than groups),
since a dense ICM is needed to distort the lobes.
Symmetrically bent, FR II galaxies may also be tracers of clusters at high
redshifts
where the richness of FR II environments is known to increase in general.  
Indeed, 
FR II's may be important in finding very high-$z$ clusters because they have 
higher luminosities and are therefore visible at larger distances than FR I's in
any flux-limited survey.  Hintzen et al. (1991) proposed looking for clusters associated
with distorted double-lobed quasars at redshifts up to 1.5.  His sample was
limited to high-power, FR II sources, around which he found suggestive
overdensities within a radius of 15$^{\prime\prime}$.

In this paper, we describe a new search for moderate- and high-redshift
clusters using a sample of bent-double radio galaxies selected from the VLA $FIRST$
(Faint Images of the Radio Sky at Twenty-cm)
survey (Becker, White, \& Helfand 1995) as tracers of cluster environments.  We demonstrate that a high
efficiency ($\sim$80\%) can be achieved during the spectroscopic phase of distant
cluster identification using our combined radio and optical selection criteria, and show that clusters with redshifts up to
$z = 0.85$ are easily detected.  In \S2, we describe the process we have 
followed to extract a bent double sample from the radio images and then
present (\S3.1) the optical imaging observations from which we selected
candidate clusters for follow-up spectroscopy.  In \S3.2, we describe the
Keck slit-mask spectroscopic observations of ten fields which led to the 
detection of at least eight clusters with redshifts ranging from 0.33 to 0.85.
Section 4 describes the radio source properties and presents a detailed discussion
of the velocity dispersions and the richness of the clusters in which they are
found.  We conclude with a summary of our results  (\S5).

We use $H_{\circ} = 50$ km s$^{-1}$ Mpc$^{-1}$ and $q_{\circ} = 0.5$ throughout
this paper.

\section{Sample}
Our sample of radio galaxies was selected from the $\it{FIRST}$ survey 
currently
being undertaken at the VLA using the B configuration.  The survey
is projected to cover the 10,000 deg$^2$ of the North Galactic Cap plus a
supplementary region between $-10^{\circ}$ and $+2^{\circ}$ Dec in the range
 21.5$<$R.A.$<$3.5 h.  The peak
flux density threshold is 1.0 mJy at 20 cm, the angular resolution is 
5$^{\prime\prime}$, and the astrometric accuracy for all detected sources is
$<1^{\prime\prime}$.  In
the $\sim$ 3000 deg$^2$ region surveyed as of the April 1997 catalog release, 
$FIRST$ has detected approximately
270,000 radio sources of which $\sim$ 32,000 are double or multiple 
sources with separations $< 60^{\prime\prime}$.  From these, we have selected
384 sources which exhibit the bent double morphology.  The peak
and integrated flux densities at 20 cm of these sources are typically 
a few mJy and
10 - 100 mJy, respectively.

The sample was selected by visually inspecting grayscale plots of
$\it{FIRST}$
sources with more than one component within a circle of radius 60$^{\prime\prime}$. 
The radio sources were examined independently by ELB, 
DJH, and RHB.  The lobes needed to be clearly bent and 'dog-leg'
FR II sources were avoided.  If two out of three of the examiners selected a source,
it was included in the sample.  Objects that were chosen by only one of us
were re-examined, and a few of these were also added to the sample.
The sample contains a large percentage of FR I radio 
galaxies, as they are more frequently found to be distorted than are FR II's 
which most often 
appear as collinear classical doubles.

Our sample of 384 objects consists of 50\% FR I sources, 25\% FR II sources, and 25\% FR I/II (sources with intermediate 
morphologies).  
Thirty-one (8\%) of these sources are within 5 arcmin of the center of
an Abell 
cluster.  We selected and classified these before knowing of their cluster associations
and found that, of these 31, we classified 80$\%$ as FR I, 10$\%$ as FR II, and 10$\%$ as FR 
I/II.  The FR I sources are easily recognizable by $FIRST$ to $z > 1$
and probably provide us with the best opportunities of finding new moderate to
high-redshift clusters.  The FR II galaxies may also be found in such 
environments and, given their higher mean radio powers, are detectable to
even higher redshifts.
Some of the sources we classified as FR I/II and II may be FR Is that have 
had extended emission resolved out by $FIRST$ and/or had their morphologies altered by
1/$(1+z)^4$ surface brightness dimming.  However, the high surface brightness and small
angular extents of the lobe hot spots which characterize FR IIs make it unlikely that an FR II
would be mistaken as an FR I.  Therefore, even after considering classification errors,
the majority of our sources are of type FR I.

Of the 384 objects, 130 were imaged in the R-band at the KPNO 2.1m and
4m telescopes between March 1995 and April 1997.  The specific 
fields for which images
were obtained were chosen following a number of different criteria.  Sources
within 5$^{\prime}$ of known clusters (31 Abell, 1 PDCS) were generally not
imaged as data already exist for them.  The majority of fields observed
contained only faint galaxies or were blank on the DSS (Digitized Sky Survey),
a selection criterion designed to be biased toward high-$z$ clusters.
We observed all sources that serendipitously lie in ROSAT pointings (except
for those in Abell clusters).  We were, of course, always constrained by the
dates observations were scheduled.  Also, in the earlier stages of the project, 
fewer sources were available
in our sample, leading to less selectivity in the fields observed (e.g., those
with brighter objects on the DSS were sometimes included).  

The selection criteria for the imaging follow-up to date are not possible to quantify.
In a future paper, observations of a clearly defined, complete subsample of
bent doubles will be
presented in order to study the relationships between source classification
(FR I or II), bending angle, radio power, and richness of the source environment.
Here, we seek simply to establish the utility of $FIRST$ bent-double radio 
galaxies as tracers of clusters at intermediate to high redshift.
We present ten fields in this paper that were chosen from the 100 imaged at 
KPNO as of December 1996.  All ten revealed only faint galaxies or were blank on the DSS.  
They were selected for
multi-object spectroscopy because the KPNO images revealed obvious 
over-densities of faint galaxies in the vicinity of the radio source.

\section{Observations and Reductions}
\subsection{Images}
Most of the optical follow-up images were obtained at the KPNO 2.1m with the 1024$\times$1024 T1KA CCD, 
giving a field of view of 5$\times$5 arcmin.  The remaining images were obtained at 
the KPNO 4m outfitted with a 2048$\times$2048 T2KB which yields a field of view
of approximately 16$\times$16 arcmin.  A log of the imaging observations for the
current sample is presented in Table 1.  The fields were observed in the R-band
in order to go as deeply as possible -- high-$z$ galaxies are relatively 
bright in the
R-band and the instrumental throughput is maximized at the wavelengths covered by this filter.
We used the KPNO Harris-R filter which was designed to match the Kron-Cousins
system.

The images were corrected for bias and flatfield effects in a standard 
manner using IRAF \footnote
{IRAF is distributed by the National Optical Astronomy Observatories, which
are operated by the Association of Universities for Research in Astronomy,
Inc., under cooperative agreement with the National Science Foundation.} (Image
Reduction and Analysis Facility).  
Typically, three observations were made of each field.  The frames were
aligned and averaged with cosmic rays rejected.

Instrumental, ``total'' magnitudes (see \S4.3 for more details) were measured using FOCAS (the Faint Object 
Classification and Analysis System, Valdes 1983).
These magnitudes were put on the standard Cousins-R scale using
observations of Landolt (1992) standards in all but three cases; for 1119+3750,
1234+4012, and 1418+3211, KPNO
spectrophotometric standards were utilized.  For these three cases, 
magnitudes at the appropriate effective wavelength (Bessel 1979) were used
to convert the instrumental magnitudes to the Cousins-R values.  Since some
of the observations were taken during nights that were not perfectly 
clear throughout,
only one or two standards were used in the calibration, typically observed
within two hours of the object during which time the sky had
changed little.  Despite imperfect weather, we believe the photometry is fairly good.  
The ten fields described here were observed during four
observing runs: 11/96 (2 nights, 5 of the fields), 3/96 (2 nights, 3 fields),
1/96 (1 field), and 3/95 (1 field).  The zero points derived for the two
nights in 11/96 were consistent to 0.05 mag.; those for the two nights in
3/96 agree to 0.03 mag.  The field observed in 3/95 (1338+4100) was later 
photometrically calibrated during a queue-scheduled observing run (R. Shuping, 
private comm.) and the resultant magnitude agreed to 0.01 mag. with that
derived using a standard star from the 3/95 observation.  The field
observed in 1/96 (1209+2848) was also calibrated during the queue run.  As a check
of photometric stability, we also measured magnitudes on each of the three
separate frames taken for each bent-double (BD) field.  The rms errors derived from the
magnitude of the host galaxy on the
three separate frames for each of the fields range from 0.01 -- 0.07, with a mean
of 0.04; these are listed in Table 1.  An additional $\sim$0.05 -- 0.10 mag. error
can be assumed because of surface brightness dimming (see \S4.3).

Previously measured KPNO atmospheric extinction corrections
were employed and the magnitude adjustments
were always very small ($< 0.02 $ mag.) because standards were observed
at airmasses very similar to those of the target frames.  Previously measured
color coefficients for the KPNO 2.1m with the T1KA CCD were used, as we observed through only one filter;
an elliptical galaxy spectrum was assumed for the radio galaxies, with
observed (V - R) colors as a function of redshift taken from Coleman, Wu, and Weedman 
(1980).  The color terms were small, increasing the galaxy brightnesses
by an average of 0.04 mag.  Corrections for Galactic extinction were taken
from the HI maps of Burstein \& Heiles (1982).  These were neglible for most of
our sources since all but one of them are at high Galactic latitudes 
($|b| > 30^{\circ}$), the sole exception being 0730+2328 at $b = 19^{\circ}$.

\subsection{Spectroscopy}
Two nights of Keck II 10m time in December 1996 were used to collect
spectra of cluster member candidates with LRIS (Low Resolution Imaging
Spectrometer, Oke et al. 1995).
We selected sixteen fields out of the 100 imaged to that point as candidates
for multi-slit mask spectroscopy and designed a mask for each.  Although one of
these two nights was lost to weather, we were able to observe five of
the fields during this run.  During an additional night in
February 1997, we obtained spectroscopy for five more fields.
The Keck fields were selected irrespective of whether the
bent radio galaxy contained therein was an FR I, II, or I/II; the primary
criterion was an obvious over-density of faint galaxies around the radio 
source on the KPNO images.

In preparation for the multi-slit masks, FOCAS was run on
astrometrically calibrated KPNO frames in order to classify objects as galaxies or stars.  
Galaxies to be observed through the masks were selected to be generally close
in brightness to the radio counterpart or fainter.  
The UCSCLRIS package (Drew Phillips, private comm.) was employed to design
the masks in IRAF.
The slit lengths were chosen to
be long enough to enable sky subtraction on both sides of the object
in the slit, and position angles were
chosen to maximize the number of galaxies on the mask.  The masks contained an
average of 22 slits over a span of up to 6 arcmin.

Observations were obtained at the Keck II 10m with LRIS 
, using a 300 lines/mm grating and the GG495 filter.  The wavelength
coverage was approximately 4950-9200 $\rm{\AA}$ for the objects at the center of the 
masks, while off-axis objects had reduced coverage.  The slit widths were 1$^{\prime
\prime}$ and the dispersion was 
2.45 $\rm{\AA}$/pixel leading to a spectral resolution of approximately 10 $\rm{\AA}$.  In addition to the slit-mask spectroscopic observations,
each field was briefly imaged without a filter; these images are used
for display purposes only in this paper.  A log of the observations is presented in Table 2.

The spectroscopic fields were reduced following standard IRAF procedures.
Each field was corrected for a bias frame and overscan region as well as
an internal halogen flat taken through the corresponding mask.  
If there was only one exposure available for
a particular field, 'cosmicrays' was run to clean the field of cosmic rays.
If two exposures were available, they were averaged after scaling and
weighting by the exposure and offsetting by the mode, and cosmic rays were rejected.
The spectra were wavelength-calibrated
using comparison spectra of Hg+Kr, Ne, and Ar taken through the appropriate
slit on the corresponding mask.  
The flux-calibration was achieved using the reduced and wavelength-calibrated 
longslit observations of standard stars --
Feige 110 for the December data and BD332642 for the February data.  Extinction
values measured at the CFHT on Mauna Kea were employed.

Initially, redshifts were approximated for each galaxy by identifying a few
obvious lines, typically the Ca II H+K absorption features and/or the
[O II] 3727, H$_{\beta}$ 4861, [O III] 4959,5007 emission lines.  The values
we present were determined by performing a Fourier cross-correlation 
between the objects'
spectra and template spectra using FXCOR in IRAF.  The elliptical galaxy
spectra were correlated with a template of M32 shifted to the rest frame.
Emission line spectra were correlated with a starburst galaxy template from
Kinney et al. (1996).  The errors in the redshifts are computed in FXCOR and
are based on the fitted peak height and the antisymmetric noise, or ``r-value''
(Tonry \& Davis 1979).  The errors are
typically 100 km s$^{-1}$ for the elliptical galaxies and 50 km s$^{-1}$ for
those objects with emission lines.
Overlays of the radio source contours onto
the direct Keck images (except for 0730+2328 which uses an image from KPNO) 
are presented in Figure 1(a-j).
Spectra for the optical identifications of the
radio sources are presented in Figure 2;  no spectrum is presented for 
0221-0202 because no obvious optical counterpart for the bent double was identified (see \S4.3).  Two spectra
are displayed for 0320-0049, one for the optical counterpart of the radio bright 
central component of the bent double, which is shown to be a quasar at $z = 
0.95$, and one for the optical
object coincident with fainter radio emission farther from the radio core which
has a spectrum consistent with that of an elliptical galaxy at $z = 0.57$.  
One spectrum is presented for each of the eight
remaining fields; most of the sources have normal elliptical galaxy
counterparts.

\section{Analysis/Results}
\subsection{Radio Galaxies}
Properties of the radio galaxies, including position, 20 cm flux density, redshift, power,
linear size, opening angle, and FR ratio, are listed in Table 3.  
The fluxes were taken from the NRAO/VLA Sky Survey (NVSS) catalog (Condon
et al. 1996), since the high
resolution of the $FIRST$ survey may resolve out some of the flux for extended 
sources. The integrated 20 cm flux densities range from 7 to 80 
mJy; the
largest discrepancy between the NVSS and $FIRST$ flux densities was $\sim25\%$.
We were unable to secure a redshift for the radio source in 0221-0202, and
0320-0049 had radio emission from optical components at very different redshifts.  The redshifts for the remaining bent-double radio galaxies range
from 0.336 -- 0.840.  All of the sources have large opening angles excepting
0819+2522 which is an NAT (or possibly a WAT viewed in projection).  The FR ratio was determined by examining the 
$\it{FIRST}$ radio
maps and applying the criteria of Fanaroff \& Riley (1974).  Edge-brightened
sources are FR II, and edge darkened sources are FR I.  We called borderline
cases 'FR I/II'.  Of the ten sources, we classify six as FR I, three
as FR I/II, and only one as FR II.
The borderline cases 1234+4012 and 1338+2931 have the largest discrepancies between the
NVSS and $FIRST$ flux densities (differences of 18\% and 24\%, respectively), 
suggesting that extended emission is being resolved out in the $FIRST$ images 
and that they are 
likely FR I sources (since FR I's have diffuse emission from their lobes).  Fanaroff \& Riley (1974) found a break in the 
powers between FR I and II sources at $\rm{2~\times~10^{25}~W~Hz^{-1}sr^{-1}}$
at 178 MHz, corresponding to $\rm{5~\times~10^{25}~W~Hz^{-1}}$ at 1440 MHz,
assuming P $\propto \nu^{-\alpha}$ and $\alpha = 0.8$.  The sources in our
sample generally have powers near the FR break, as is typical of WAT
galaxies (O'Donoghue et al. 1993).

\subsection{Cluster Redshifts and Velocity Dispersions}
Histograms of the redshift distribution of galaxies in each of the ten
fields are presented in Figure 3.  The bins are 0.02 wide in $z$, 
corresponding to 4000 km s$^{-1}$ at $z = 0.5$. Seven of the fields show clear peaks
in redshift space that include the radio galaxy (indicated with **).  
For 0221-0202, no radio galaxy counterpart has been identified, although there
is a clear peak in the $z$ distribution at $z=0.33$.   The source 0320-0049 
has a quasar at the center of the bent 
double (marked with **), and a radio galaxy (marked with *) which
is coincident with a peak in the $z$ distribution that does not include the
quasar. The redshift peaks for
both 0221-0202 and 0320-0049 may result from groups or clusters that
we recognized on the KPNO images but which are unassociated with the 
bent-double radio sources.
In 0730+2328, there are four galaxies at $z \approx 0.85$, including the 
radio galaxy; combining this with the richness for this field, derived in \S4.3,
we believe that this represents a significant detection of
a cluster.  The redshifts for individual
cluster member galaxies are listed in Table 4 and those of non-members are 
presented in Table 5.

The line-of-sight velocity dispersions are calculated for the cluster-member galaxies in our
fields using $\sigma_{\|}=\sqrt{(N-1)^{-1}\sum_{i=1}^{N}\Delta v_i^2}$, where
$\Delta v_i=c(z_{i}-\bar z)/(1+\bar z)$;
they are listed in Table 4.  The 68\% confidence uncertainties are 
calculated following Danese,
De Zotti, and di Tullio (1980), and include errors due to the small number
of member galaxies per field and the measurement errors.  We find a range in
the dispersion values from approximately 300 to 1100 km s$^{-1}$.  The large 
velocity dispersions in some of the
fields may be explained in three ways:  1) we have included unrelated field 
galaxies as cluster members; 
2) the clusters are not virialized, having 
undergone recent cluster-cluster (or subcluster) merging; or 
3) the clusters have large masses.
To address the first
possibility, we 'clean' our cluster membership
list using the method of Yahil and Vidal (1977).  The galaxy with the largest
velocity offset from the cluster mean is excluded, $\sigma_{\|}$ is recalculated,
and the new offset for the questionable galaxy is determined.  If the new 
offset is greater than 3$\sigma_{\|}$, the galaxy is removed from the member
list.  This process is repeated with the next galaxy until no further members
are
rejected.  Applying this to the galaxies listed in Table 4 would have a
significant effect on one field,
and marginally modify our results on two others.   In 1119+3750, if galaxy \#17 (3.42 $\sigma_{\|}$
offset) and then \#8 (3.66 $\sigma_{\|}$ offset) are
removed, we obtain a new velocity dispersion of 322 km s$^{-1}$ (instead of
709 km s$^{-1}$), 
and a new
mean redshift of 0.4868.  The radio galaxy's velocity offset from the mean 
is lowered to
-425 km s$^{-1}$ from -706 km s$^{-1}$.  One galaxy each in 1234+4012 and 1338+2931 is a
borderline case.  If \#4 is removed from 1234+4012, its offset from the new
mean (0.4875) is 3.00 times the new $\sigma_{\|}$ (350 km s$^{-1}$, instead of
532 km s$^{-1}$).  The radio
galaxy's velocity offset from the mean lowers to 20 km s$^{-1}$ from 182 km
 s$^{-1}$.  Removing galaxy
\#9 from 1338+2931 gives a new mean redshift of $z=0.6387$ and $\sigma_{\|}=483$
km s$^{-1}$ (instead of 640 km s$^{-1}$);
\#9's offset is 3.07$\sigma_{\|}$.  In this case, the radio galaxy's velocity offset actually rises
to 934 km s$^{-1}$ from 787 km s$^{-1}$.  Since this procedure may remove galaxies involved in a cluster
or subcluster merger, or otherwise related galaxies, as well as unrelated field galaxies, it is not necessarily
desirable.

High-resolution velocity histograms, with 200 km s$^{-1}$ bins, are presented 
in Figure 4 (a), where velocities are shown relative to the cluster mean.  These histograms
are not simple, smooth Gaussian distributions centered on the cluster mean velocity; however,
the distributions shown $are$ typical of known clusters and are to be expected when velocities
are measured for so few cluster members.  For comparison, in Figure 4 (b), we present 
velocity histograms for the Coma cluster.  To make these plots analogous to
those for the BD clusters, ten
of the thirty brightest galaxies in Coma were randomly selected, velocity dispersions were
calculated, and the histograms were constructed.  Again, we do not see Gaussian distributions
centered on the cluster mean velocity, and what appear to be clumps of outlying galaxies
are sometimes present.  The velocity dispersions calculated from these simulations range from
673 -- 1251 km s$^{-1}$.  The true velocity dispersion of the galaxies in the main Coma cluster
(not including a group of galaxies which appears to be merging with the main cluster), 
is 1082 km s$^{-1}$, and the overall distribution is Gaussian (Colless \& Dunn 1996).
The large uncertainties listed for our BD cluster velocity dispersions in Table 4 attest to the 
substantial uncertainty of our measurements.

It is not possible to distinguish between  the other two explanations for
large velocity dispersions
without X-ray observations of each field or many more measured redshifts that could reveal substructure in
the clusters.  In fact, the clusters with large $\sigma_{\|}$ may well be both massive and dynamically
active.  In a ROSAT study of the X-ray emitting
gas in clusters containing WATs, G\'omez et al. (1997) found that the direction of the bending of
the radio lobes often corresponds to the orientation of the X-ray isophotes, 
suggesting that merging cluster gas is providing the ram pressure necessary
for bending the lobes of radio galaxies lying at the centers of cluster 
potential wells.   If, indeed, WATs are formed as a result of 
cluster-cluster merging, we would expect the environments
of WATs to be not completely virialized and their associated clusters to 
have large velocity
dispersions.  If a radio galaxy identified with a cD galaxy has a significant
peculiar motion relative to the cluster, we also expect that the cluster is
incompletely virialized.  In a study of the kinematics of dense clusters of
galaxies, Zabludoff et al. (1993) found that ``systems with speeding cD's are
probably a guide to substructure in dynamically evolving systems.''  The
radio galaxy may be at the kinematic center of a subclump involved in the 
merger of poor clusters or groups and thus would have a significant peculiar
velocity relative to the cluster as a whole.  Merritt (1985) suggested that
cD's are formed in such mergers (cf. Zabludoff et al. 1993).
We see a range in the offset of the velocity of the radio galaxy with
respect to the mean cluster velocity of 70 to 1220 km s$^{-1}$ ($\Delta v$,
Table 4).  Several of the radio galaxies are travelling with velocities
of a few hundred km s$^{-1}$ with respect to the cluster mean. This is faster than
expected from simulations of cD galaxies in virialized clusters 
(Malumuth 1989) but slower than the $\sim 1000$ km s$^{-1}$
required to bend the lobes in existing models (O'Donoghue et al. 1993), so
these clusters may well represent merging systems. 

If the velocity dispersions $\it{are}$ approximately representative of 
the cluster
masses, we calculate masses in the range $1\times10^{14}~\rm{M_{\odot}}$
(1209+2848) to $8\times10^{14}~\rm{M_{\odot}}$ (0819+2522) within a radius
of approximately one megaparsec.  At the high end of this range, these would
be some of the highest-$z$ massive clusters known.  
In the following section, we go beyond those galaxies for which we have
directly measured redshifts to examine the richness of the fields surrounding
each radio source.

\subsection{Richness}
We measure the richness in the KPNO images of the fields surrounding 
each source in two different ways, following the
example of H\&L91 in the first case, and that of Allington-Smith
et al. (1993) and Z97 in the second.  The H\&L91 sample included approximately
50 radio galaxies with $z \approx 0.5$, and with a range of radio powers, although
only 13 were classified as FR I.  The Z97 sample includes a larger range in
redshifts with $z < 0.5$, and all of the sources have high powers ($\approx
20\%$ are FR I's, and these are almost exclusively at low-z in her sample).  The H\&L91 method uses N$_{0.5}$, the number of galaxies above the background 
galaxy counts within a radius of 0.5 Mpc of the
radio galaxy and within the magnitude range m$_{R,rg}$ to m$_{R,rg}$
+ 3, where m$_{R,rg}$ is the apparent isophotal R-magnitude of the radio 
galaxy measured to a surface brightness limit of 27.5 mag./arcsec$^2$ (their
optical observations are similar to ours -- 15 - 30 min. total exposures in
the R-band at a 2.2m telescope, limiting magnitude 23).
This type of richness measure is similar to an Abell richness parameter which includes
galaxies with magnitudes in the range m$_3$ to m$_3$ + 2, where m$_3$ is the
magnitude of the third brightest cluster member.  
The modification to m$_{R,rg}$ + 3  
was made because the third brightest cluster
member is difficult to recognize with increasing redshift, due to the superposition of
field galaxies along the line-of-sight. 
The H\&L richness measurement assumes that the radio galaxy is the first
ranked member of the cluster, which is not always the case.  If it is not, we measure
the H\&L parameter twice:  once using the radio galaxy's magnitude as the starting point,
and again using that of the brightest cluster galaxy.
The advantage of using the H\&L technique to measure
the richness is that it does not assume a form for the luminosity function 
of the
cluster as a function of redshift, as does the richness measure B, the correlation statistic used by Yates, Miller, and Peacock (1989) and Yee and
Green (1984, 1987) among others, although it does assume that the evolution
for all cluster members is similar.

The method introduced by Allington-Smith et al. (1993) and also used by Z97
uses N$_{0.5}^{-19}$, the number of galaxies within a radius of 0.5 Mpc
from the radio galaxy and down to M$_V = -19$ at the redshift of the radio
source.  The advantage of this technique is that it will sample the same range
of absolute magnitudes in each cluster, as opposed to H\&L's technique 
which will sample different ranges since the absolute magnitudes of radio
galaxies differ (by as much as $\sim1$ magnitude in M$_V$ for our sample).
However, the quantity N$_{0.5}^{-19}$ is dependent on K-corrections and
color assumptions for the cluster galaxies.

We used FOCAS to find the positions of all objects detected at levels 
at least 3$\sigma$ above
the sky, classify them as galaxies or 
stars, split merged objects
into their components, and calculate FOCAS ``total'' magnitudes.  The 
area used for determining the ``total'' magnitude is found by increasing
the area used in calculating the isophotal magnitude by a factor of two.
The ``total'' magnitudes reach a surface brightness limit of $\approx 27.5$ 
mag/arcsec$^2$ for our data.  Because of 1/$(1+z)^4$ surface brightness 
dimming, an isophotal limit of 27.5 mag/arcsec$^2$ at $z = 0$ translates
to a limit of 26.3 mag/arcsec$^2$ at $z = 0.33$ and 24.8 mag/arcsec$^2$ at
$z = 0.85$ (the redshift extremes of our clusters).  Tests that measured the magnitudes at different isophotal limits
tell us that the true total magnitudes of the galaxies in our clusters are
0.05 - 0.1 mag. brighter than our measured ``total'' magnitues, but this will
not affect our richness measurements.  To improve sky subtraction, we increased
the distance of the sky buffer (the distance from the edge of an object's isophote
to the edge of the annulus where sky is measured) from the FOCAS default value to
be large enough that magnitudes no longer changed as the buffer distance was increased
(this turned out to be 20 pixels).

To determine N$_{0.5}$ (\`a la H\&L), we counted all galaxies within a radius 
of 0.5 Mpc of the radio galaxy, defined by the redshift of the radio galaxy
and using $q_{\circ} = 0.5$ and $H_{\circ} = 50~\rm{km~s^{-1}~Mpc^{-1}}$,
within the magnitude range m$_{R,rg}$ to m$_{R,rg}$ + 3.
Due to the difficulty in measuring magnitudes in exactly the same way as other
authors, we decided to measure background (field) galaxies directly from our
cluster images, even though there will be some contamination from cluster galaxies.
We counted galaxies in an annulus extending as far out in each cluster frame as possible
and having the same area as the aperture in which we counted cluster + field
galaxies.  Due to the size of our frames and the redshifts of the clusters,
the inner radii of the annuli were $\lesssim 1$ Mpc and so the annuli should 
still include some cluster galaxies.  This will make our richness measurements
conservative (underestimates of the true richness).  As a comparison, we also
calculated the number of galaxies expected within the appropriate magnitude range
and angular area
using the number counts from two deep surveys:  Tyson (1988), whose counts agree well
with those derived from H\&L's background fields, and Metcalfe et al. (1991), whose counts
agree with other surveys such as Steidel \& Hamilton (1993) and Couch et al. (1993).
Tyson (1988) found log(N) = 0.39R -- 4.8, where N is the number of galaxies per
magnitude per square degree and R is the apparent isophotal R-magnitude to
a surface brightness limit of 28 mag/arcsec$^2$ (which Tyson found to be
essentially equivalent to the total magnitude).  Metcalfe et al. (1991) found
log(N) = 0.373R -- 4.51, where N is the number of galaxies per 0.5 magnitude per 
square degree and R is an aperture magnitude with a correction applied to make it
a total magnitude.  We would expect the number density of galaxies derived
from field surveys
to be lower than those from our annuli as the annuli counts will include cluster
galaxies in the outskirts of the clusters.  For the ten bent-double fields, we found
the number of background galaxies expected from Tyson (1988) to be a mean of 22$\pm13\%$
$lower$ than our annulus counts, and those from Metcalfe et al. (1991) to be 32$\pm22\%$
 $higher$ than the annulus counts.

For each cluster candidate, the annulus background counts were subtracted from the raw N$_{0.5}$ counts.
We define our limiting magnitude in the
same way as H\&L:  0.5 brighter than the peak in the number vs. magnitude histogram
constructed for each frame.  A correction factor was applied if m$_{R,rg}$ + 3 was fainter than our
limiting magnitude (m$_R \approx 23$ for our 15 minute exposures and 
m$_R \approx 23.5$ for our 30 minute exposures). 
  In these cases galaxies
were counted down to the limiting magnitude only and background counts were
determined in that same range.  The correction factor was determined assuming
a Schechter (1976) luminosity function with M$^{*}_{V} = -21.9$ and $\alpha = -1.25$,
and was defined as $f_{c} = \phi(m_{R,rg}+3) / \phi(m_{lim})$.  This value
was multiplied by the raw N$_{0.5}$, the counts above the background. 
In addition to calculating N$_{0.5}$ centered on the position of the radio
galaxy and using the magnitude of the radio galaxy as the starting m$_{R}$,
we followed the same procedure with the brightest cluster galaxy's position
and magnitude (when the radio galaxy and BCG were not one and the same).  Only galaxies
that were spectroscopically confirmed cluster members were considered when
choosing the BCG.  

In order to determine a richness value using the same intrinsic luminosity 
range for cluster members, we also measured 
N$_{0.5}^{-19}$.  To transform apparent R-magnitudes to absolute
V-magnitudes, an elliptical galaxy spectrum was assumed with a rest frame color
(V -- R) = 0.9, and K-corrections were taken from Coleman, Wu, and Weedman
(1980).  No evolution correction was applied.  Galaxy counts were then
made within an 0.5 Mpc radius centered on the radio galaxy, including all
those with M$_V$ brighter than $-19$.  When this was fainter than the limiting
magnitude of the observation, a Schechter correction was applied in a fashion 
similar to that described above.  The correction factors for N$_{0.5}^{-19}$ 
were
usually greater than those for N$_{0.5}$, since M$_V = -19$ at the redshift
of the radio galaxy was usually fainter than m$_{\rm{R,rg}}$ + 3 or m$_{\rm{R,bcg}}$ +
3. 

Table 6 lists the results.  Columns (3) and (4) list the apparent R-magnitudes
of the radio galaxies and brightest cluster galaxies, respectively.  When
the radio galaxy is the BCG, no value is listed for m$_{R,bcg}$.  Column (5) lists
the rms errors derived from each field's three separate frames.  Total errors may
be approximately 0.05 -- 0.10 mag. higher because of surface brightness dimming.
These errors have a negligible effect on our richness measurements. Columns (6) and (7)
list the absolute V-magnitudes of the radio galaxies and BCGs, respectively.  Columns
(8), (10), and (12) list the corrected values of N$_{0.5}^{-19}$, 
N$_{0.5,rg}$, and N$_{0.5,bcg}$, respectively, with Poisson errors from the 
cluster and background counts, and each column followed by
$f_c$, the correction factor that was applied to give the corrected counts.
In the last column, an estimate of the Abell richness class is given for 
each of the clusters.  This is approximated using Z97's conversion
of $\rm{N_{Abell} = 2.7(N_{0.5}^{-19})^{0.9}}$, and Abell's (1958) conversion
between $\rm{N_{Abell}}$ and richness class.  Before using this relation, 
we corrected to $q_{\circ} = 0$, as was used by Z97
-- this affects the sampling radius in which galaxies are counted, and
decreases the richness values by $\approx 15\%$ (as shown by H\&L91).
Using H\&L91's correlation between
N$_{0.5}$ and Abell class leads to slightly higher richnesses.

We find that all of the clusters associated with bent doubles in this sample
fall in the Abell richness class range of 0 to 2, typical of the majority
of clusters in Abell's catalog of rich clusters.  The mean $\rm{N_{Abell}}$ in Abell, Corwin,
\& Olowin's 1989 catalog is 60, which is Abell
richness class 1, or $\rm{N_{0.5}^{-19} = 31}$.
The fields around two of our
bent doubles, 0221-0202 and 0320-0049, contain poor clusters or groups.
The system in 0320-0049 does not appear to be associated 
with the bent double which itself
may have a richer associated cluster at higher redshift.  
The optical i.d. for
0221-0202 is uncertain, and it is not at all clear that the radio source is
associated with the surrounding group or cluster.  There is a peak in the redshift
distribution for this field, and the BCG was selected from
that peak for use in determining the richness.  
The object at the core of the
bent double in 0320-0049 has a quasar spectrum, with a redshift of 0.95, 
and is behind the group of galaxies.  However, a galaxy included in
the group located close to the core of the bent double, does show some radio
emission (although this object would most probably not have been selected if not
for the radio emission from the quasar).  It is the magnitude and redshift
of this galaxy that was used in determining the richness for this field.
Finally, the field around 1234+4012 contains two groups or clusters , one associated 
with the radio galaxy at $z = 0.49$, and the other at $z = 0.31$.
Therefore, N$_{0.5,rg}$ probably gives a better approximation of
the richness than N$_{0.5}^{-19}$ since N$_{0.5,rg}$ excludes many of the brighter,
foreground $z \approx 0.3$ galaxies.  It is this richness (N$_{0.5,rg}$) 
that is used in 
approximating the Abell class for this cluster.

Excluding 0221-0202, 0320-0049, and 1234+4012 for the reasons cited above,
and 0730+2328 because it has such a large correction factor, the mean is
N$_{0.5}^{-19} = 28\pm8$ for $q_{\circ} = 0.5$ or 24$\pm7$ corrected to $q_{\circ} = 0$ to be consistent with 
Z97, which is Abell's class 0 using Z97's
conversion.  This is to be compared with Z97's mean value of 10.2 for all
of the radio-selected groups in her sample (or 14.3 for FR I's and 8.7 for
FR II's).  
Excluding 0221-0202, 0320-0049, and 0730+2328, the mean is N$_{0.5,rg} = 19\pm5$
(Abell's class 1, using H\&L91's conversion), compared
with H\&L91's mean value of 9.7 for all radio sources in their sample.

These results show that our sources are found in richer environments than radio sources in
general.  This difference is most apparent when compared with the environments
of FR II sources, but is still marked even when compared with Z97's FR I
subsample (which have a mix of unbent and bent morphologies).
However, remembering that the sample of ten sources in this paper was drawn
from a much larger sample of bent radio galaxies after examining optical
follow-up images and seeing obvious over-densities of galaxies, we cannot
yet conclude that, in general, bent radio galaxies lie in richer environments
than other radio sources because
of our bias towards rich environments in the optical selection.  Nonetheless,
we have demonstrated that our combined selection criteria produce an
efficient method for finding clusters at $z\ge0.5$.
In a future paper (Blanton et al. 1999), we will
present a larger, lower-$z$, unbiased, and complete (area and magnitude limited)
subsample of our $FIRST$ bent double sample which will determine whether or not
bent doubles, as a whole, are found in richer environments than other radio sources.

\section{Conclusions}
We have shown that bent-double radio sources selected from the $FIRST$ survey
are effective tracers of cluster environments at moderate to high redshifts.
Of the ten fields studied, eight have clear evidence of being associated
with clusters in the redshift range $0.33 < z < 0.85$, and we cannot rule out that
the remaining two may be associated with clusters at higher redshifts.  The clusters
display a range of line-of-sight velocity dispersions, from approximately 300 to 1100 
km s$^{-1}$, and a richness range of Abell class approximately 0 to 2.  On the
upper end of these ranges, we may be sampling some of the highest-$z$ massive
clusters known, and/or we may be seeing systems with significant substructure
that have recently undergone mergers.  
CDM models with $\Omega = 1$ predict strong evolution in the number density
of massive clusters with redshift.
The discovery of just a few high-redshift
massive clusters within the area covered by the $\it{FIRST}$ survey would
help put firm upper limits on $\Omega$, possibly ruling out $\Omega = 1$, 
CDM models (Bahcall, Fan, \& Cen 1997).  For now, it is difficult
to say whether the clusters we have presented are massive, have substructure, 
or both.   X-ray observations will distinguish amongst these possibilities.

The authors thank Drew Phillips for providing the software 
package (UCSCLRIS) that was used in designing the slit masks, Ralph Shuping for 
obtaining calibration data during a queue-scheduled observing run at the KPNO
0.9m, Frank Valdes for answering FOCAS questions, and the referee for
helpful comments.
Part of the work reported here was done at the Institute of Geophysics and
Planetary Physics, under the auspices of the U.S. Department of Energy by
Lawrence Livermore National Laboratory under contract No.~W-7405-Eng-48.
We acknowledge use of the NASA/IPAC Extragalactic Database (NED) which is
operated by the Jet Propulsion Laboratory, Caltech, under contract with the
National Aeronautics and Space Administration.
The $\it{FIRST}$ project is supported by grants from the National 
Geographic Society, the National Science Foundation (AST-94-19906), 
NASA (NAG5-6035), NATO, IGPP, Columbia University, and Sun Microsystems.
This is Columbia Astrophysics Laboratory Contribution No. 654.

\newpage
\noindent
Fig. 1.(a.--j.)  Overlays of the radio source contours onto
the direct Keck images (except for 0730+2328, Fig. 1c., which uses an 
image from KPNO).
Numbers marking the galaxies correspond to
those listed in Table 4 (large, bold-face numbers marking cluster members) and Table
5 (small numbers for non-members).  Tables 4 and 5 are discussed in \S4.2.
The lowest radio contour for each overlay is listed in its caption; 
the contours are then 1, 2, 4, 8, 16, 32, and 64 times the lowest level.
\bigskip

\noindent
Fig. 1.a.  0221$-$0202, $z = 0.33$, 7 confirmed members, richness $<$ Abell 0, total flux density = 7.3 mJy, lowest contour = 0.425 mJy.
\bigskip

\noindent
Fig. 1.b.  0320$-$0049, $z = 0.57$, 5 confirmed members, richness $<$ Abell 0, total flux density = 67.0 mJy, lowest contour = 0.5 mJy.
\bigskip

\noindent
Fig. 1.c.  0730+2328, $z = 0.85$, 4 confirmed members, richness = Abell 2, total flux density = 9.1 mJy, lowest contour = 0.35 mJy.
\bigskip

\noindent
Fig. 1.d.  0819+2522, $z = 0.56$, 9 confirmed members, richness = Abell 0, total flux density = 65.1 mJy, lowest contour = 0.7 mJy.
\bigskip

\noindent
Fig. 1.e.  0910+3841, $z = 0.51$, 10 confirmed members, richness = Abell 1, total flux density = 78.8 mJy, lowest contour = 0.57 mJy.
\bigskip

\noindent
Fig. 1.f.  1119+4216, $z = 0.48$, 10 confirmed members, richness = Abell 1, total flux density = 31.3 mJy, lowest contour = 0.45 mJy.
\bigskip

\noindent
Fig. 1.g.  1209+2848, $z = 0.34$, 8 confirmed members, richness = Abell 0, total flux density = 15.1 mJy, lowest contour = 0.4 mJy.
\bigskip

\noindent
Fig. 1.h.  1234+4012, $z = 0.49$, 6 confirmed members, richness = Abell 1, total flux density = 45.3 mJy, lowest contour = 0.5 mJy.
\bigskip

\noindent
Fig. 1.i.  1338+4100, $z = 0.64$, 11 confirmed members, richness = Abell 1, total flux density = 29.1 mJy, lowest contour = 0.45 mJy.
\bigskip

\noindent
Fig. 1.j.  1418+3211, $z = 0.36$, 15 confirmed members, richness = Abell 0, total flux density = 28.5 mJy, lowest contour = 0.35 mJy.
\bigskip

\begin{figure}
\figurenum{2}
\plotone{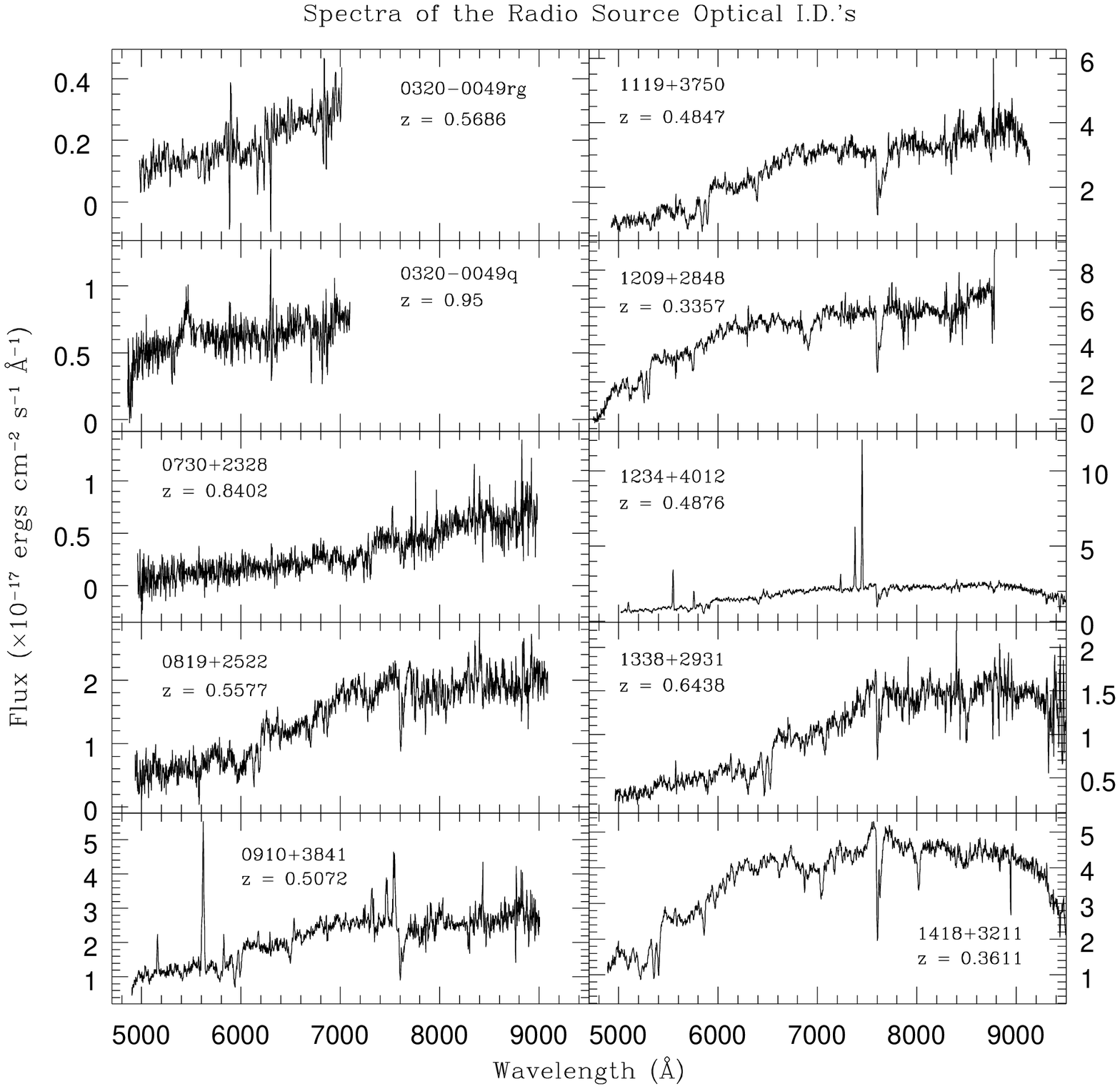}
\caption{Spectra from the Keck II LRIS of the optical identifications for the 
bent-double radio sources.  Most of the hosts are elliptical galaxies.}
\end{figure}

\begin{figure}
\figurenum{3a}
\plotone{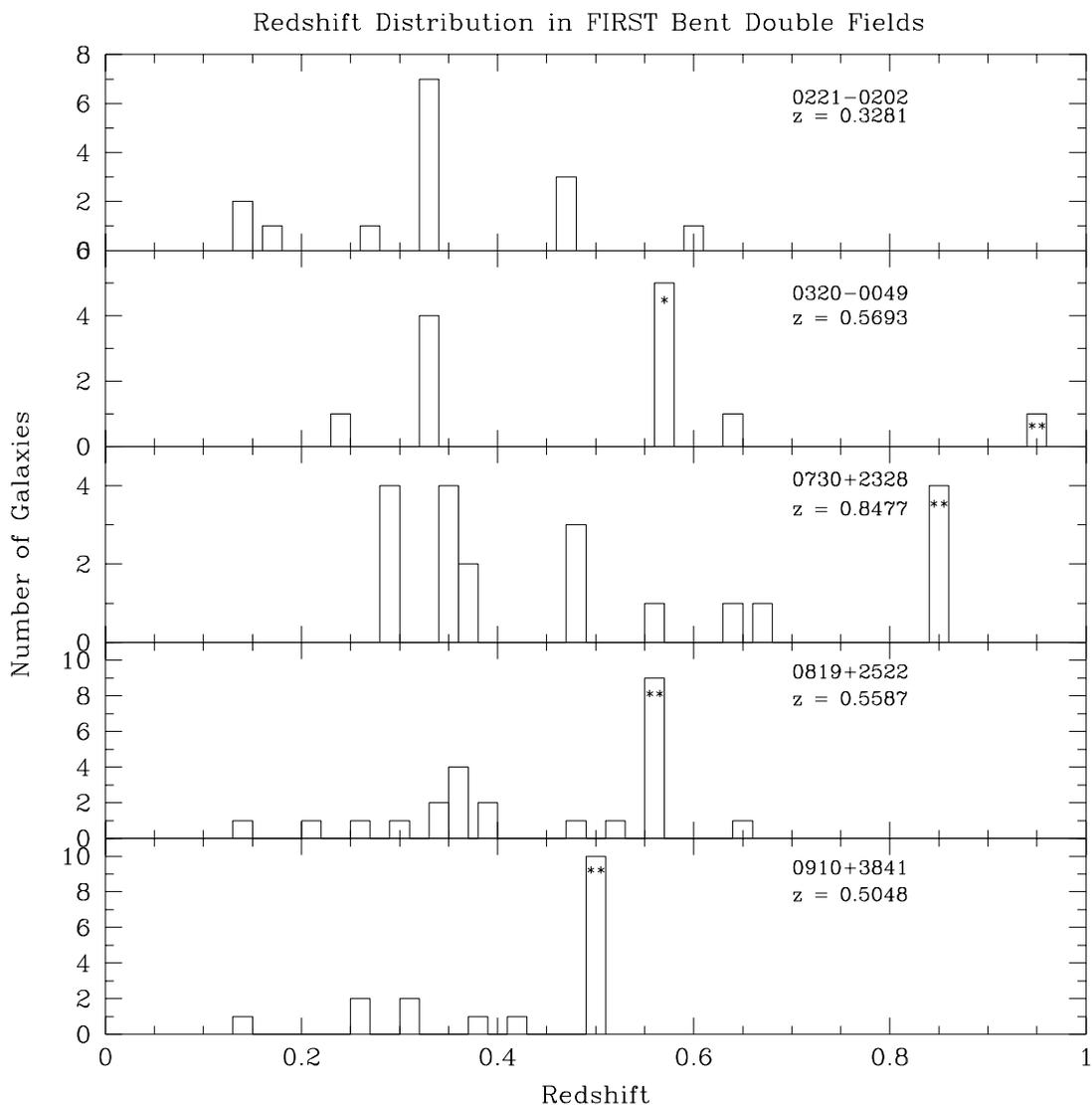}
\caption{The distribution of galaxy redshifts in the fields surrounding the
bent-double radio sources.  Seven of the fields show obvious evidence for
a cluster associated with the bent double (marked with **).  One other
(0730+2328) has four galaxies coincident with the bent double redshift; combined with richness measurements, this also seems to be a cluster.  There is
a peak in the redshift distribution in 0320-0049 which includes a radio
source (marked with *) adjacent to the bent double.  The optical i.d. to
0221-0202 is not obvious and is thus not marked on the histogram.}
\end{figure}

\begin{figure}
\figurenum{3b}
\plotone{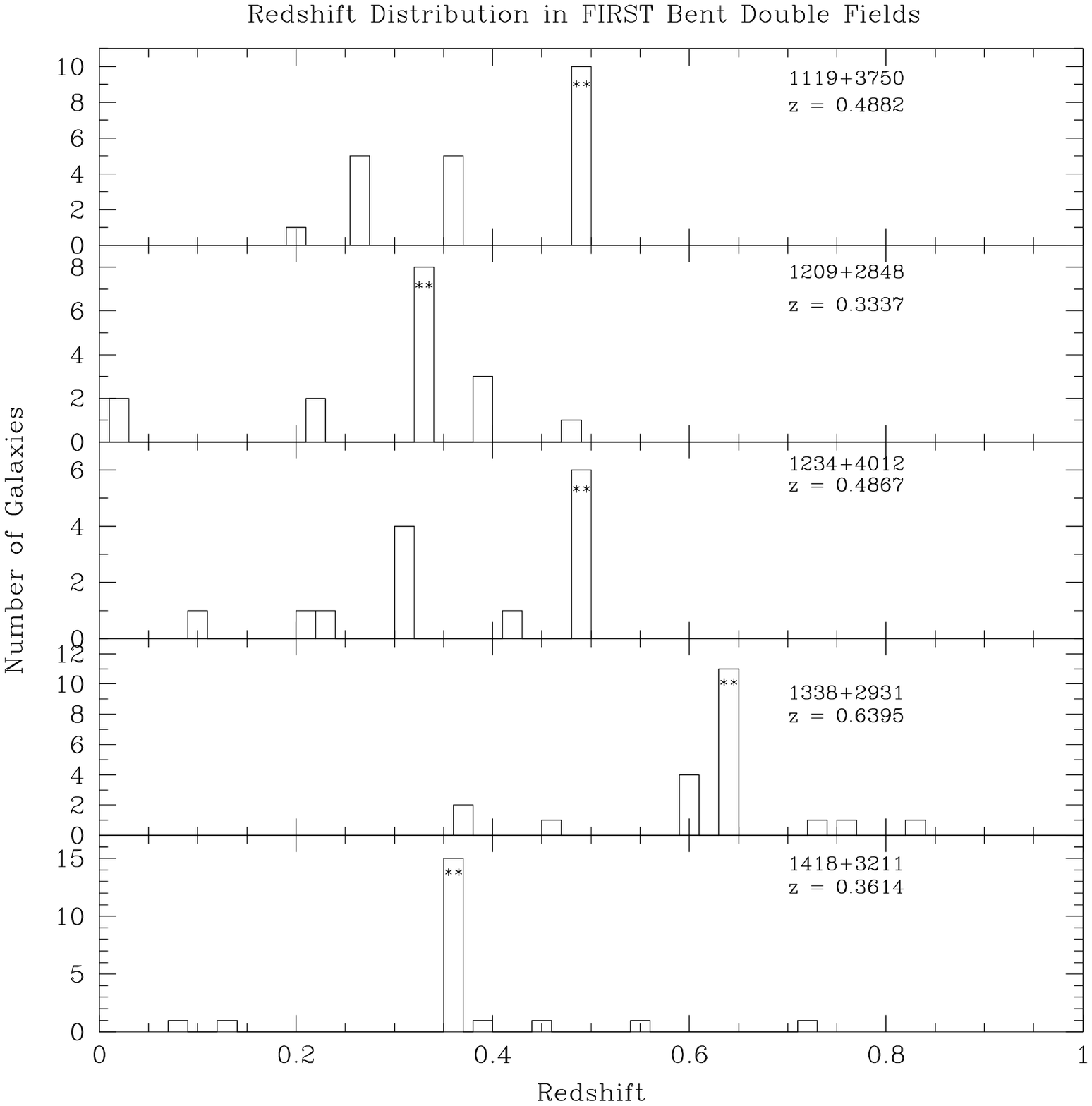}
\caption{Figure 3, continued.}
\end{figure}

\begin{figure}
\figurenum{4a}
\plotone{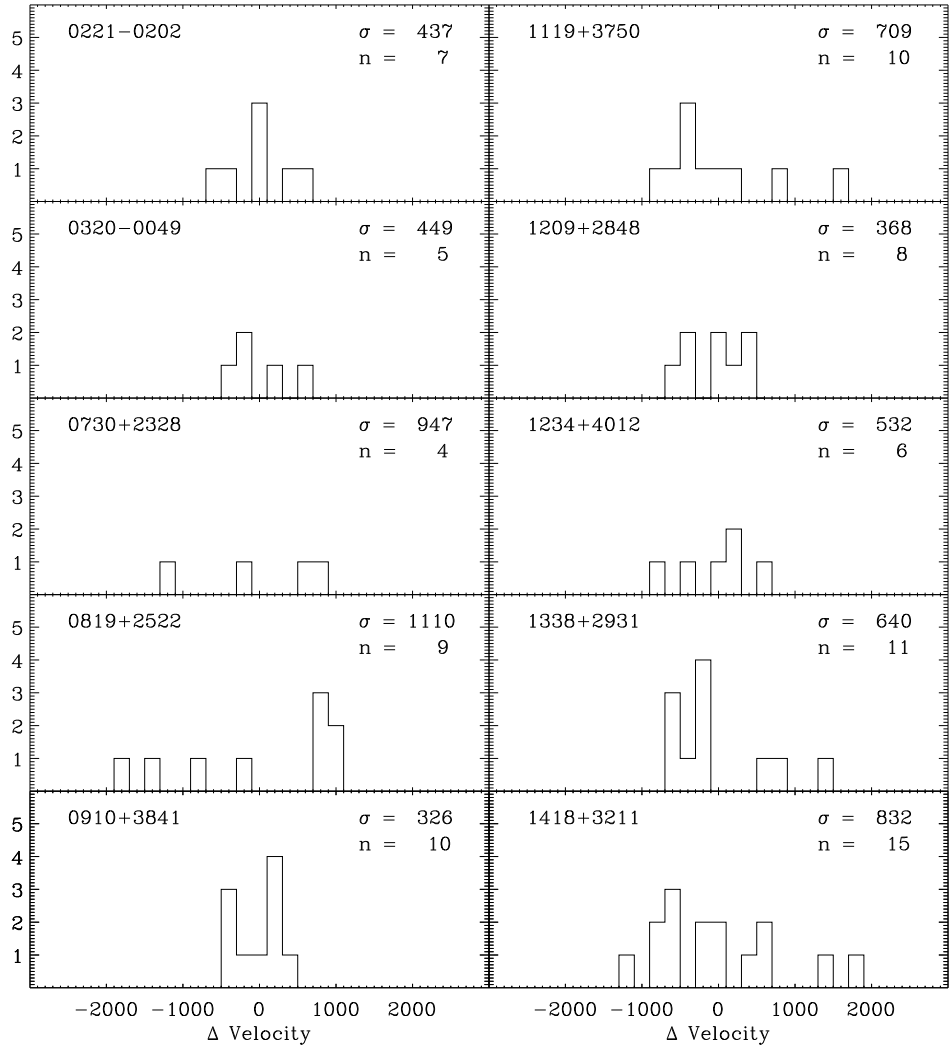}
\caption{High-resolution velocity dispersion histograms of the BD clusters.
Bins are 200 km s$^{-1}$ wide and velocities are shown relative to the cluster mean.}
\end{figure}

\begin{figure}
\figurenum{4b}
\plotone{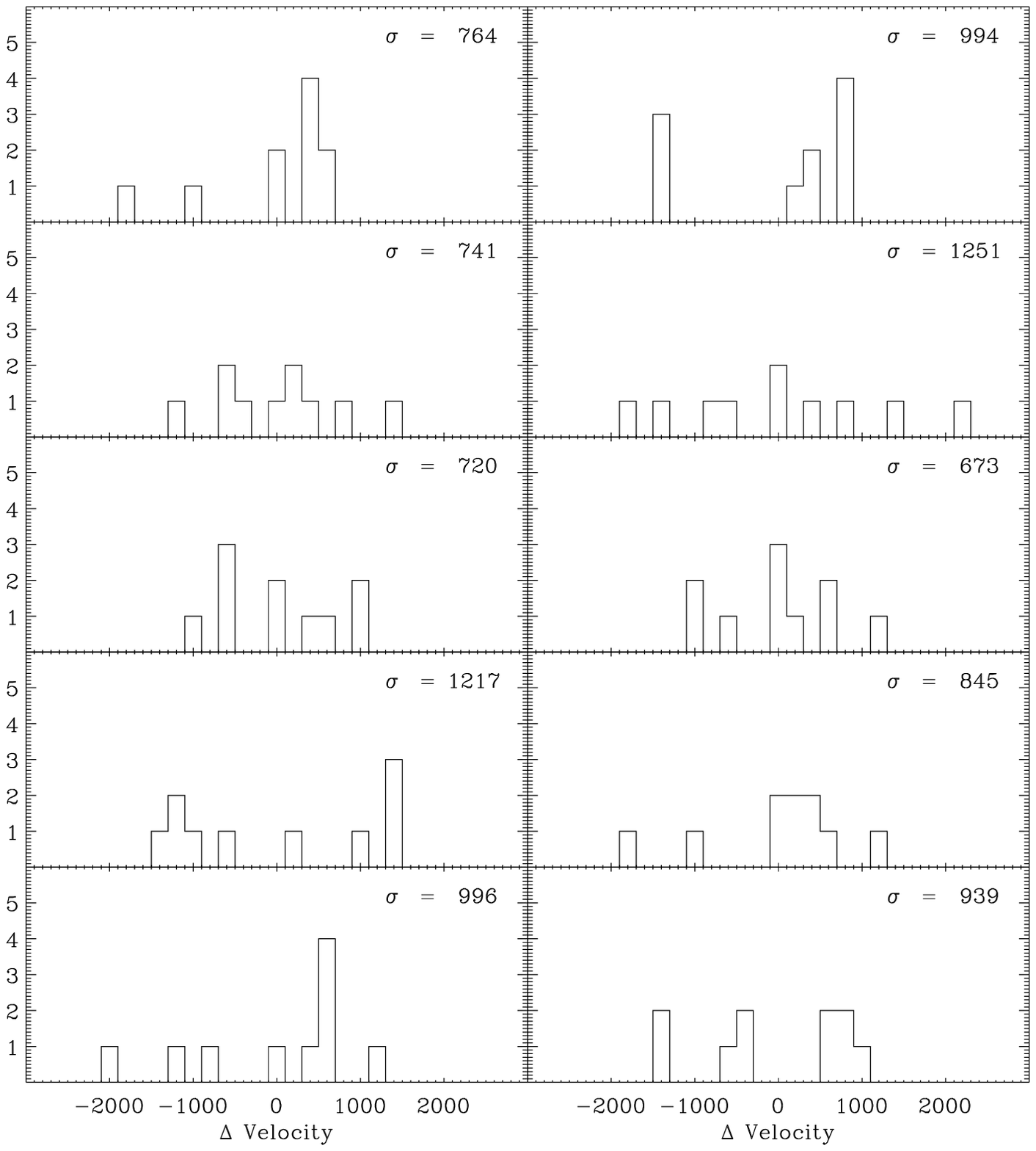}
\caption{Velocity dispersions histograms for the Coma cluster.  To make these plots 
analagous to those for the BD clusters, ten
of the thirty brightest galaxies in Coma were randomly selected, velocity dispersions were
calculated and histograms were constructed.  Again, we do not see Gaussian distributions
centered on the cluster mean velocity.}  
\end{figure}

\plotone{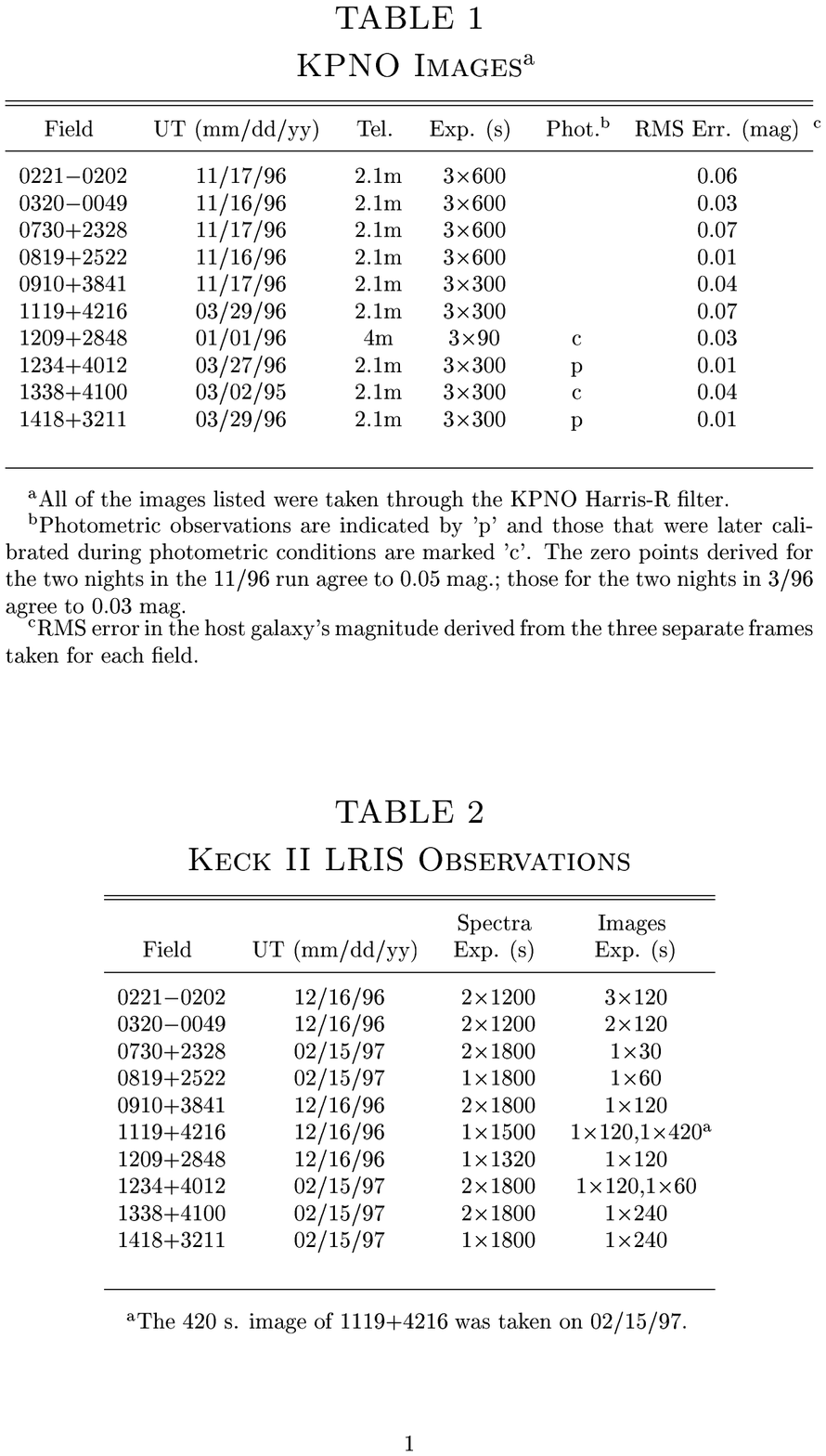}

\plotone{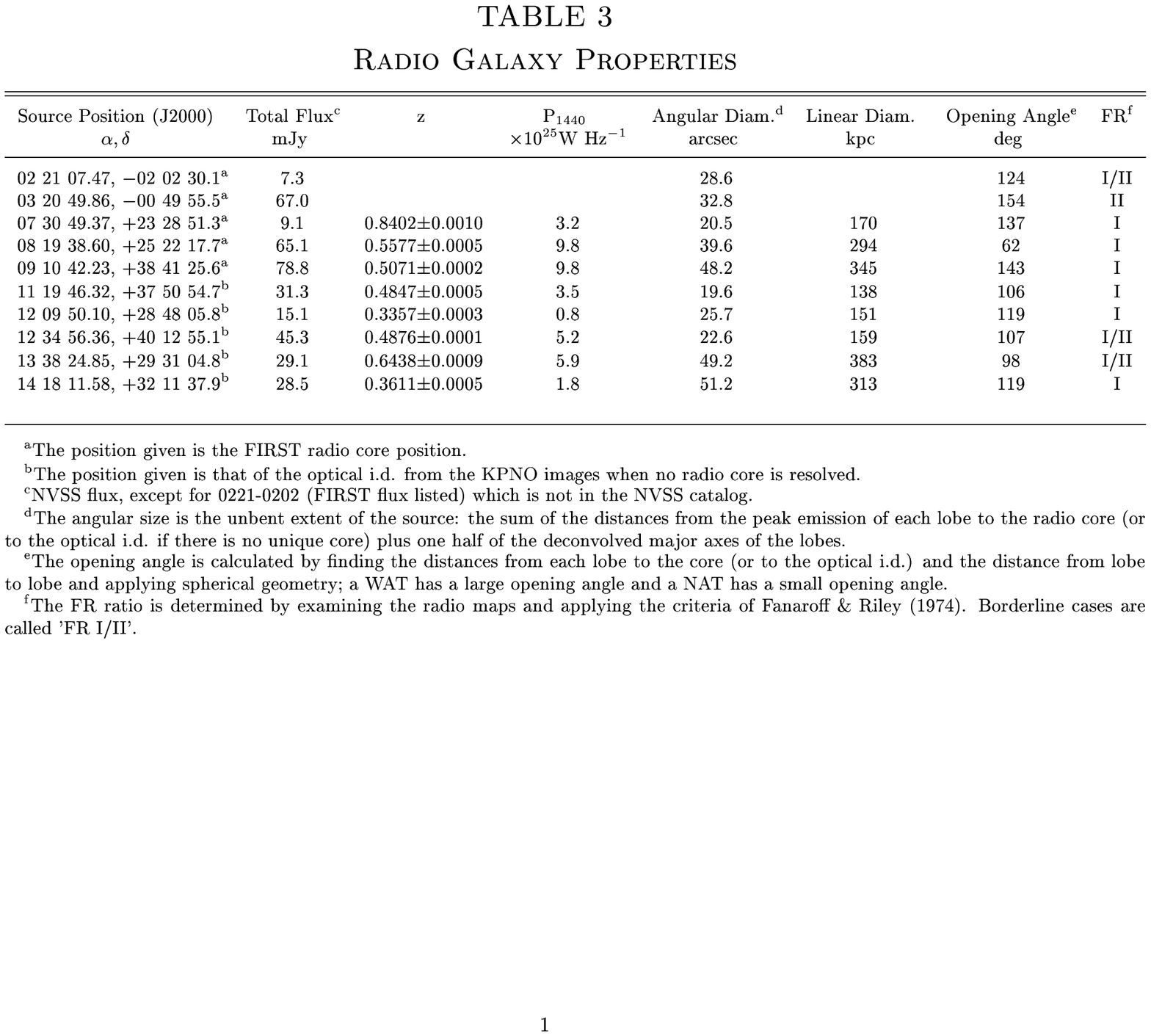}

\plotone{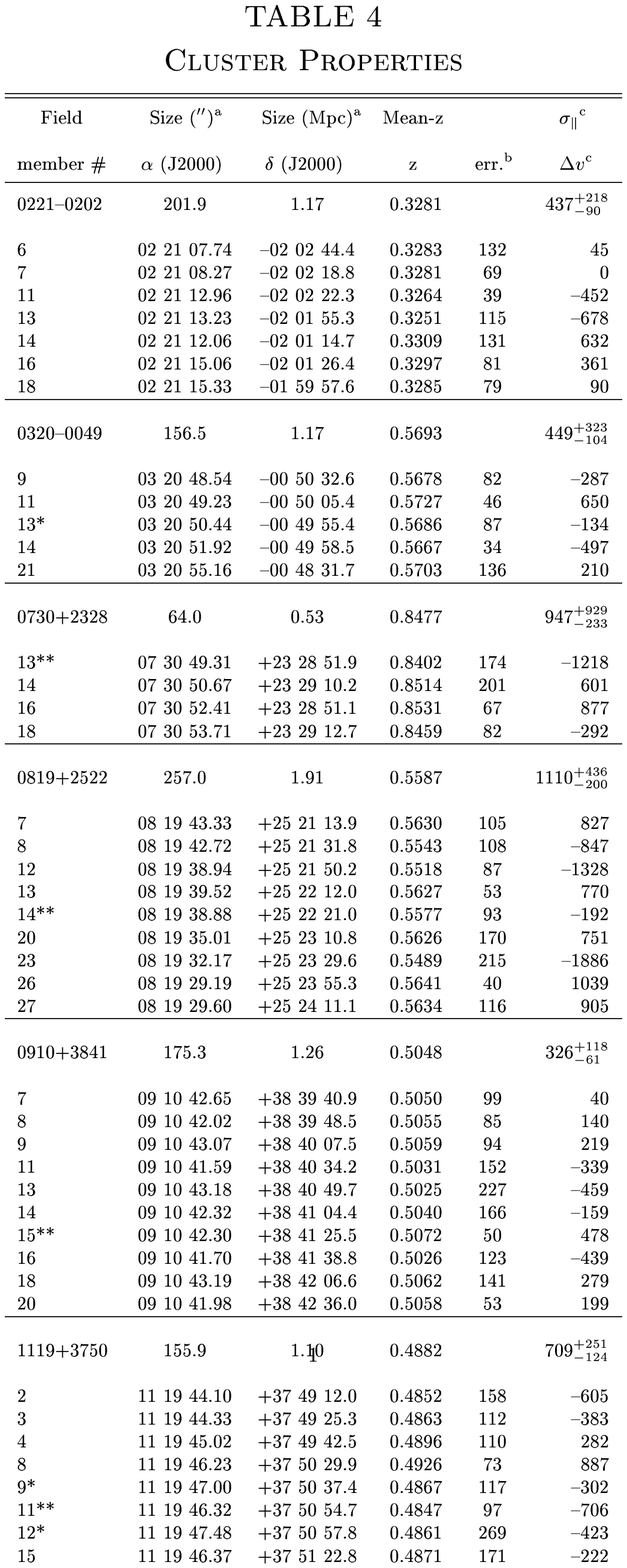}

\plotone{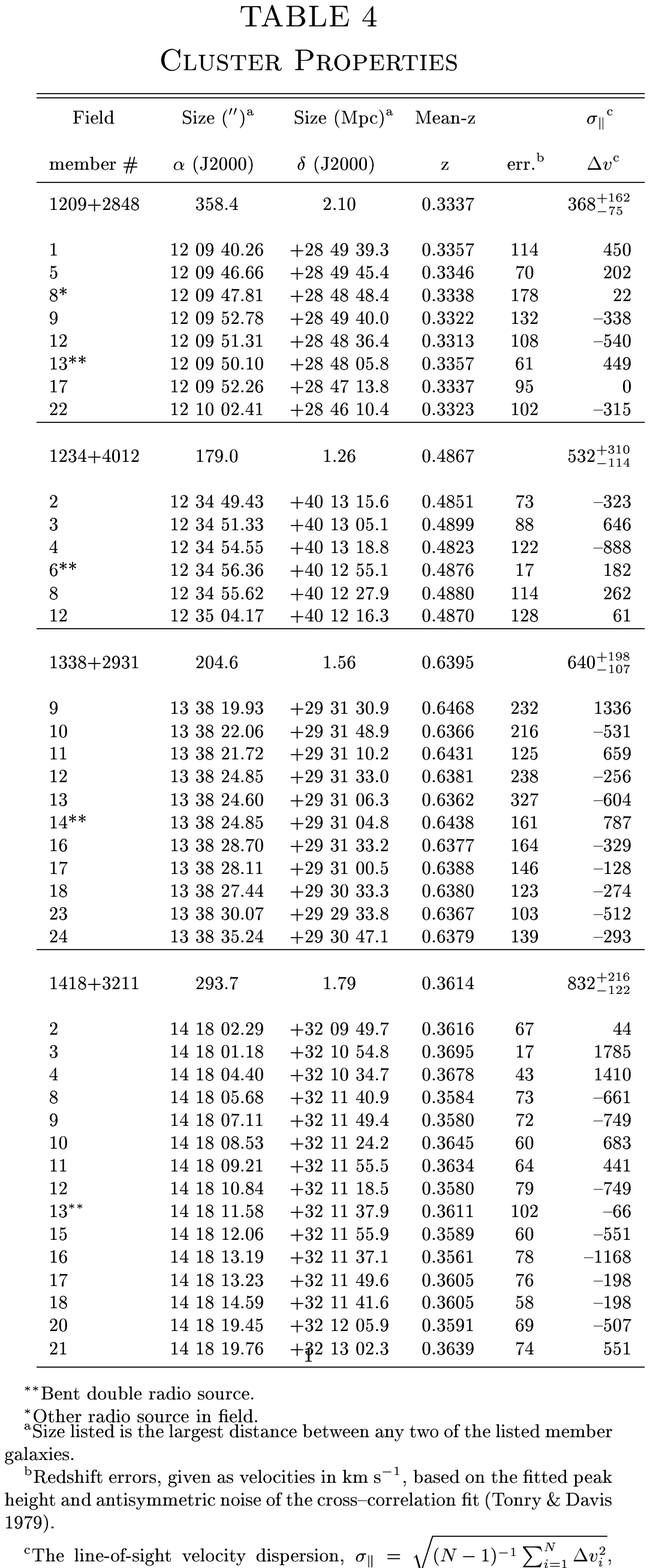}

\plotone{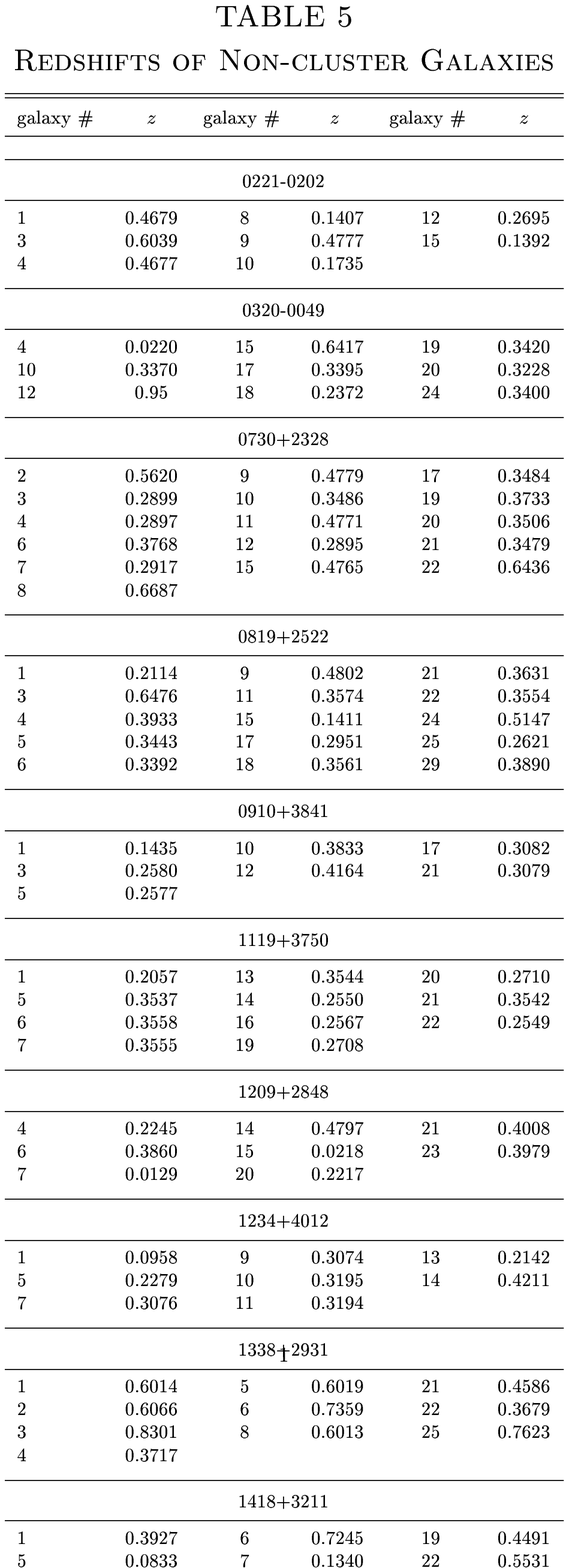}

\plotone{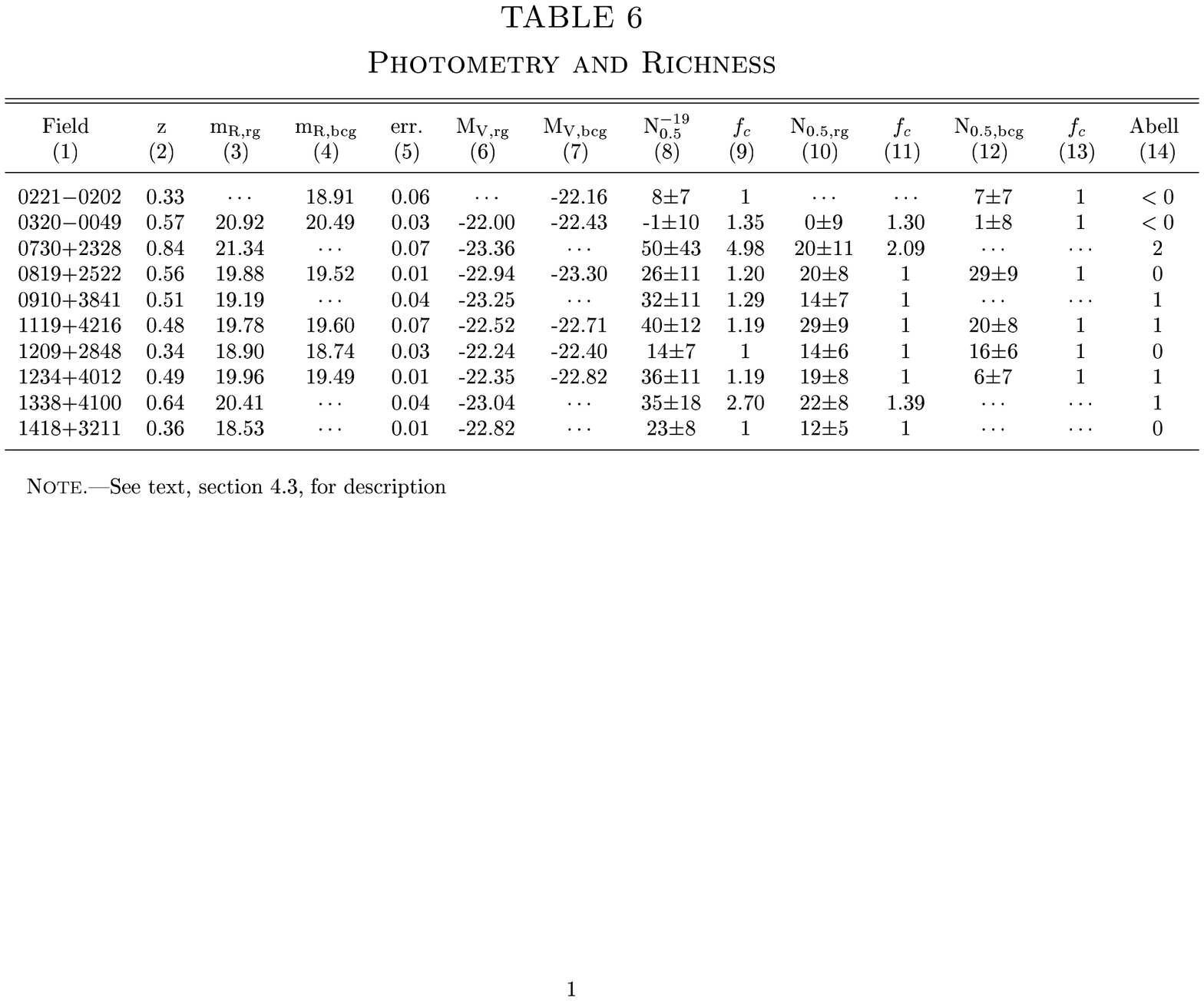}

\end{document}